\documentclass{article} 

\usepackage{microtype}
\usepackage{graphicx}
\usepackage{subcaption}
\usepackage{csquotes}
\usepackage{afterpage}
\usepackage{xcolor}

\usepackage{cuted}
\usepackage{caption}

\usepackage{amsmath}
\usepackage{amssymb}
\usepackage{mathtools}
\usepackage{amsthm}
\usepackage{xspace}
\usepackage{colortbl}
\usepackage{multirow}
\usepackage{enumitem}
\usepackage{wrapfig}
\usepackage{nicefrac}
\usepackage[font=small,labelfont=bf,justification=centering]{caption}

\usepackage{booktabs}
\usepackage{multirow}
\usepackage{graphicx}
\usepackage{xcolor}
\usepackage{pifont} 
\usepackage{threeparttable}

\usepackage{geometry}
\usepackage{tabularx}

\usepackage{amsmath}

\usepackage{iclr2025_conference,times}

\usepackage{amsmath,amsfonts,bm}

\def\eqref#1{equation~\ref{#1}}

\def\1{\bm{1}}

\DeclareMathAlphabet{\mathsfit}{\encodingdefault}{\sfdefault}{m}{sl}
\SetMathAlphabet{\mathsfit}{bold}{\encodingdefault}{\sfdefault}{bx}{n}

\usepackage{hyperref}
\usepackage{url}

\title{MM-Sonate: Multimodal Controllable Audio-Video
Generation with Zero-Shot Voice Cloning}

\author{
  Chunyu Qiang\thanks{Equal contribution} \quad
  Jun Wang\footnotemark[1] \quad
  Xiaopeng Wang \quad
  Kang Yin \quad
  Yuxin Guo \\
  Kuaishou Technology \\
  \texttt{\{qiangchunyu, wangjun06, wangxiaopeng05, yinkang03, guoyuxin07\}@kuaishou.com}
}

\iclrfinalcopy 
\begin{document}

\renewcommand{\thefootnote}{\fnsymbol{footnote}}

\maketitle

\begin{center}
    \includegraphics[width=1.0\textwidth,keepaspectratio]{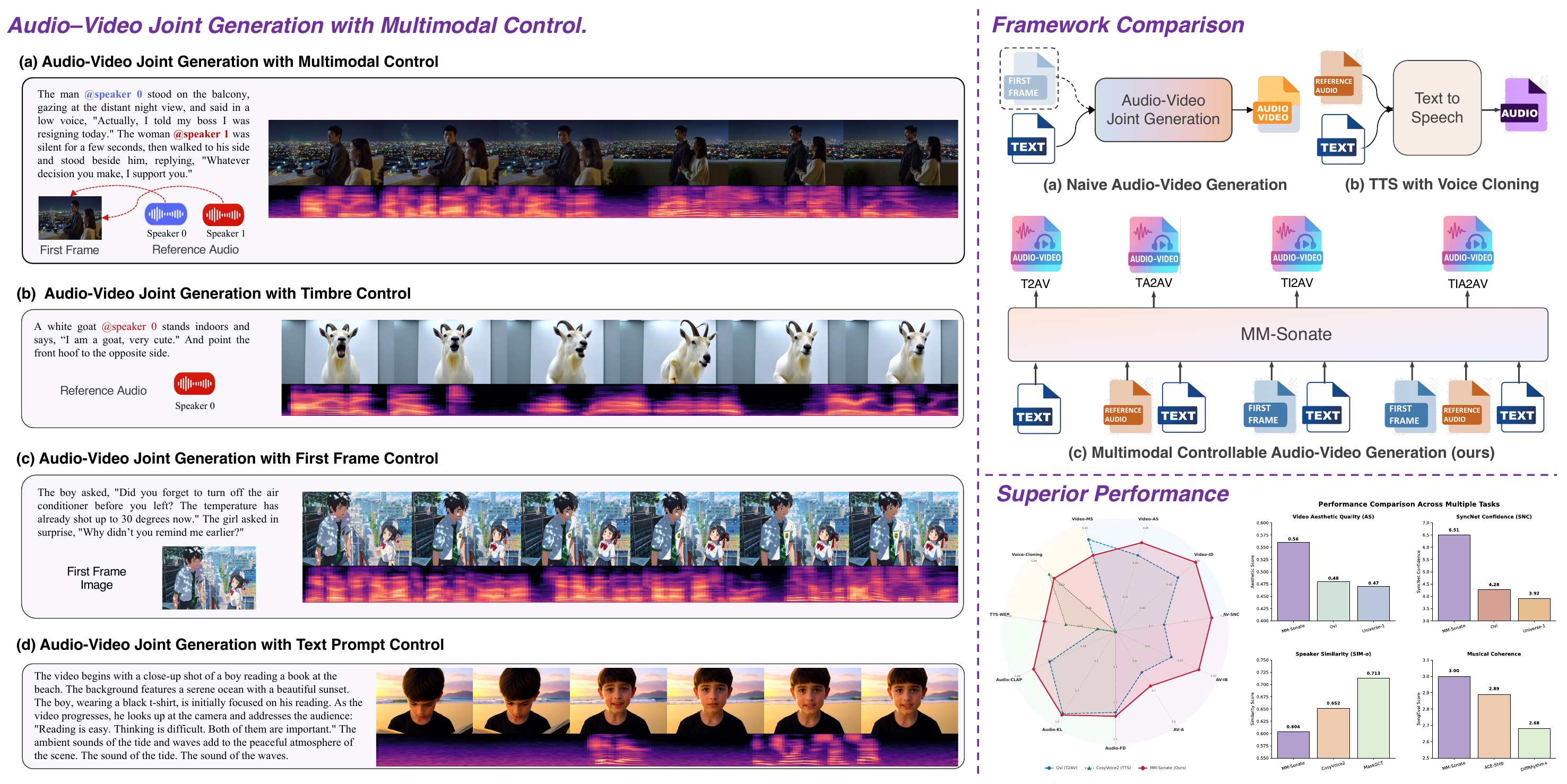}
    
    \vskip 1mm 
    
    \captionof{figure}{
    Audio–Video Joint Generation with Multimodal Control: synthesizing coherent, temporally aligned dialogue conditioned on the first-frame and distinct reference audios.
    Unlike existing frameworks, our unified architecture supports flexible combinations of multi-modal inputs for precise control.
    MM-Sonate outperforms baselines (Ovi, CosyVoice2) on 9/11 metrics.
    }
    \label{fig:teaser} 
    
    \vskip 5mm
\end{center}

\begin{abstract}
Joint audio-video generation aims to synthesize synchronized multisensory content, yet current unified models struggle with fine-grained acoustic control, particularly for identity-preserving speech. Existing approaches either suffer from temporal misalignment due to cascaded generation or lack the capability to perform zero-shot voice cloning within a joint synthesis framework. In this work, we present MM-Sonate, a multimodal flow-matching framework that unifies controllable audio-video joint generation with zero-shot voice cloning capabilities. Unlike prior works that rely on coarse semantic descriptions, MM-Sonate utilizes a unified instruction-phoneme input to enforce strict linguistic and temporal alignment. To enable zero-shot voice cloning, we introduce a timbre injection mechanism that effectively decouples speaker identity from linguistic content. Furthermore, addressing the limitations of standard classifier-free guidance in multimodal settings, we propose a noise-based negative conditioning strategy that utilizes natural noise priors to significantly enhance acoustic fidelity. Empirical evaluations demonstrate that MM-Sonate establishes new state-of-the-art performance in joint generation benchmarks, significantly outperforming baselines in lip synchronization and speech intelligibility, while achieving voice cloning fidelity comparable to specialized Text-to-Speech systems.
\end{abstract}

\section{Introduction}
\label{section:intro}
The ultimate goal of generative world simulation is to replicate the multisensory nature of physical reality, necessitating the synchronous synthesis of visual and acoustic signals. While recent text-to-video diffusion models have achieved remarkable visual fidelity, the integration of high-quality, synchronized audio remains a significant challenge. Early approaches typically treated audio generation as a post-processing step, employing cascaded systems where audio models are conditioned on pre-generated video frames~\cite{mmaudio, wang2025kling, Luo2023DiffFoleySV, wang2025audiogen, guo2025audiostory, guo2025aligned}. However, this decoupled paradigm fundamentally struggles with fine-grained temporal alignment, frequently resulting in significant asynchrony between video and audio streams.

To address the alignment problem, research has shifted toward unified architectures that generate audio and video within a single model. Pioneering works such as MM-Diffusion~\cite{ruan2023mm} and JavisDiT~\cite{liu2025javisdit} explored joint modeling, while state-of-the-art models like Ovi~\cite{low2025ovi} achieve impressive synchronization through symmetric dual-backbone designs. Despite these advances, a critical limitation persists: existing unified models lack fine-grained controllability over the acoustic modality, particularly in speech generation. For instance, while Ovi can generate semantically relevant sounds, it cannot control the specific identity (timbre) of a speaker or clone a voice from a reference clip. Consequently, current unified video generators cannot yet function as high-fidelity, personalized speech synthesis, limiting their application in scenarios requiring consistent character identity or dubbing.

In this paper, we propose MM-Sonate, a multimodal controllable framework designed for joint audio-video generation with zero-shot voice cloning capabilities. Built upon the Multi-Modal Diffusion Transformer (MM-DiT) architecture and flow matching techniques, MM-Sonate treats audio and video as coupled streams within a shared latent space. Unlike prior approaches that rely solely on text descriptions for audio control, we introduce a unified instruction-phoneme input format. This enables the model to leverage both semantic understanding for scene generation and phonetic information for precise lip synchronization. Crucially, to enable zero-shot voice cloning, we design a timbre injection mechanism that decouples speaker identity from linguistic content, allowing the model to synthesize speech in the voice of an unseen reference speaker without fine-tuning.
Furthermore, we identify that the standard Classifier-Free Guidance (CFG) strategy, which typically uses a null embedding for the unconditional branch, is suboptimal for diverse acoustic conditions. Through ablation studies, we demonstrate that using a zero vector as a negative prompt merely steers the model away from ``random voice generation'' rather than low-quality artifacts. To address this, we propose a novel inference strategy using naturally collected noise as a negative speaker embedding, which significantly enhances speech stability and speaker consistency.

To support zero-shot voice cloning capabilities, we construct a high-fidelity synthetic dataset specifically filtered for timbre consistency, alongside a large-scale pre-training corpus of 100 million aligned audio-video pairs. This data strategy empowers MM-Sonate to generalize across a diverse range of generation tasks.
Our main contributions are summarized as follows:
\begin{itemize}
\item We propose MM-Sonate, a unified flow-matching framework that achieves state-of-the-art performance in audio-video joint generation. By integrating a unified instruction-phoneme input, the model supports flexible combinations of text, image, and audio conditions.
\item We introduce the first joint generation model capable of zero-shot voice cloning. Through a dedicated timbre injection mechanism, MM-Sonate achieves speaker similarity and speech intelligibility comparable to specialized Text-to-Speech (TTS) systems, while simultaneously generating synchronized video.
\item We propose a noise-based negative conditioning for CFG. We empirically show that using natural noise as a negative prior yields superior acoustic performance compared to standard zero-vector baselines.
\item Extensive evaluations demonstrate that MM-Sonate outperforms existing unified and cascaded baselines in both objective metrics and human preference, particularly in lip synchronization precision and speech intelligibility.
\end{itemize}


{
  \renewcommand{\thefootnote}{\fnsymbol{footnote}}
  \footnotetext[1]{Equal contribution.}
  \footnotetext[2]{``Sonate'' is derived from the musical term ``Sonata'', symbolizing the model's comprehensive capabilities in audio generation.}
}

\section{Related Works}
\subsection{Audio-Video Joint Generation} 

Audio-Video Joint Generation aims to simultaneously synthesize visual frames and acoustic signals, typically conditioned on text descriptions, ensuring both semantic consistency and temporal synchronization across modalities within a unified generation process.
Pioneering efforts like MM-Diffusion~\cite{ruan2023mm} introduced this task using coupled U-Net backbones but were often constrained by small-scale datasets (e.g., roughly 10 hours), which limited their generalization capabilities. Following the architectural shift in unimodal generation, recent works have predominantly adopted Diffusion Transformers (DiTs). These approaches generally fall into two architectural paradigms: unified or coupled streams. For instance, UniForm~\cite{zhao2025uniform} employs a unified single-tower architecture to process concatenated audio-video tokens, whereas AV-DiT~\cite{wang2024av} adapts a pre-trained image DiT with lightweight adapters to handle multimodal signals.
A central challenge in this field remains precise spatio-temporal synchronization. To address this, JavisDiT~\cite{liu2025javisdit} introduces a hierarchical prior, while Ovi~\cite{low2025ovi} enforces tight coupling through a symmetric twin-backbone design. Similarly, SyncFlow~\cite{liu2024syncflow} utilizes a dual-DiT architecture with Rectified Flow Matching to enhance alignment stability. Apart from training from scratch, another line of research orchestrates pre-trained unimodal experts to leverage strong priors; MMDisCo~\cite{hayakawa2024mmdisco} uses a discriminator for cooperative guidance, while Universe-1~\cite{wang2025universe} fuses specialized models at the block level.
Despite advances in architecture and data scale—with recent works leveraging larger corpora like VGGSound~\cite{chen2020vggsound} and AudioSet~\cite{gemmeke2017audio}—most models are confined to generating general sound effects or music. The synthesis of synchronized speech alongside video remains a significant and underexplored challenge due to the requirement for fine-grained lip synchronization. 

\subsection{Voice Cloning} 
Voice Cloning aims to mimic a certain speaking style, including timbre, prosody, emotion and more nuanced features. Substantial advancements have been achieved for this task, especially from the scope of TTS (text-to-speech), in a zero-shot way. The conditioning pattern for the voice cloning task are of three general paradigms: (1) providing solely an audio prompt for in-context learning~\cite{wang2023neural,chen2024f5,qiang2024minimally,qiang2024high,wang2025emotional, wang2025m3}, where all features of the given speaker are reserved but deeply entangled and uncontrollable; (2) further leveraging a speaker embedding (or derived from a specific aspect of speaking style, e.g. emotion) to provide overall guidance for synthesis~\cite{du2024cosyvoice,cui2025glm,zhou2025indextts2,yu2025joyvoice}, which allows partial controllability within a relatively simple model design; (3) instructing with text prompt depicting desired variations of the cloned voice~\cite{ji2024controlspeech,du2024cosyvoice2,yin2025dmp}, which offers extremely high flexibility but still underexplored. Although there have been many explorations in pure speech tasks, there have been few studies on cross-domain transition (covering tasks such as music and sound) and multimodal generation in terms of voice cloning. 

\section{Preliminary}
\subsection{Multimodal Diffusion Transformer} 
Our model employs a unified conditioning architecture that accommodates diverse input modality configurations—including text, image, and audio—in arbitrary combinations.  This design enables the model to handle various generation scenarios, such as T2VA (text-to-video-and-audio), TI2VA (text-and-image-to-video-and-audio), TA2VA (text-and-audio-to-video-and-audio), and TIA2VA (text-image-audio-to-video-and-audio), within a single framework.
The proposed approach builds upon the MM-DiT architecture introduced in SD3\cite{sd3}.  During training, each input modality is processed by dedicated encoders, with absent modalities represented by trainable placeholder tokens.  After independent encoding, all modality-specific features are projected into a shared embedding space to facilitate cross-modal integration.

\subsection{Flow Matching} 
\label{flow matching}
Our training framework leverages conditional flow matching \cite{lipman2022flow} as the generative modeling objective \cite{tong2023improving}. The approach learns a parameterized velocity field $v_\theta(t,C,x)$ conditioned on inputs $C$ (such as encoded text or video representations) that characterizes the transformation dynamics at each timestep $t$, where $\theta$ denotes the trainable network weights. We adopt the Optimal Transport(OT) path to construct the flow, defining the ground-truth vector field as $u_t(x|x_1) = x_1 - (1-t)x_0$. This learned field enables the transformation of an initial Gaussian noise sample $x_0$ into the target audio latent $x_1$ through numerical integration via an ODE solver across the temporal interval $t\in[0, 1]$. Here, $p$ indicates the conditional probability trajectory and $q$ represents the empirical data distribution.

\begin{equation}
\label{e:cfm}
    \mathcal{L}_{CFM}(\theta) = \mathbb{E}_{t,q,p} \big\| v_\theta(t, C, tx_1+(1-t)x_0) - u(t, tx_1+(1-t)x_0) \big\|^2,
\end{equation}

At inference time, we apply the Euler integration scheme with a timestep increment of 0.05 to numerically solve the learned velocity field $v_\theta(t,C,x)$, progressively transforming random Gaussian noise into the desired audio latent representation.

\subsection{Audio-Video Latent Representation}

\textbf{Latent Audio Encoder-Decoder.} The latent audio codec extends our prior SecoustiCodec framework\cite{qiang2025secousticodec,qiang2025vq,qiang2024learning}, inheriting its core architecture while introducing key modifications to optimize audio reconstruction quality. The codec employs a Mel-VAE structure comprising three primary components: a mel-spectrogram encoder, a mel-spectrogram decoder, and a discriminator. The audio encoder processes input waveforms sampled at 44.1 kHz and compresses them into continuous latent embeddings at a substantially reduced temporal rate, achieving high compression ratios while preserving perceptual quality. The VAE architecture enables the model to learn continuous and complete distributions in the latent space, significantly enhancing audio representation capabilities and reconstruction fidelity

\textbf{Latent Video Encoder-Decoder.} For visual representation, we adopt the 3D causal temporal encoder from CogVideoX\cite{yang2024cogvideox} as our video codec backbone. The video encoder processes input frames at a resolution of $H \times W$ and temporal length $T$, compressing them into spatiotemporal latent embeddings with dimensionality $h \times w \times t$, where $h \ll H$, $w \ll W$, and $t \ll T$. The encoder applies progressive spatial-temporal downsampling, achieving a total compression ratio of $(H/h) \times (W/w) \times (T/t)$ to produce compact latent representations. Correspondingly, the video decoder mirrors the encoder architecture with transposed convolutions, reconstructing high-fidelity video frames from the latent embeddings. Similar to the audio codec, the video VAE learns a continuous latent distribution, facilitating smooth interpolation and stable training dynamics in the flow matching framework.

\section{MM-Sonate}
\label{section:stage1}

\begin{figure*}[t]
 \centering
 \includegraphics[width=\linewidth]{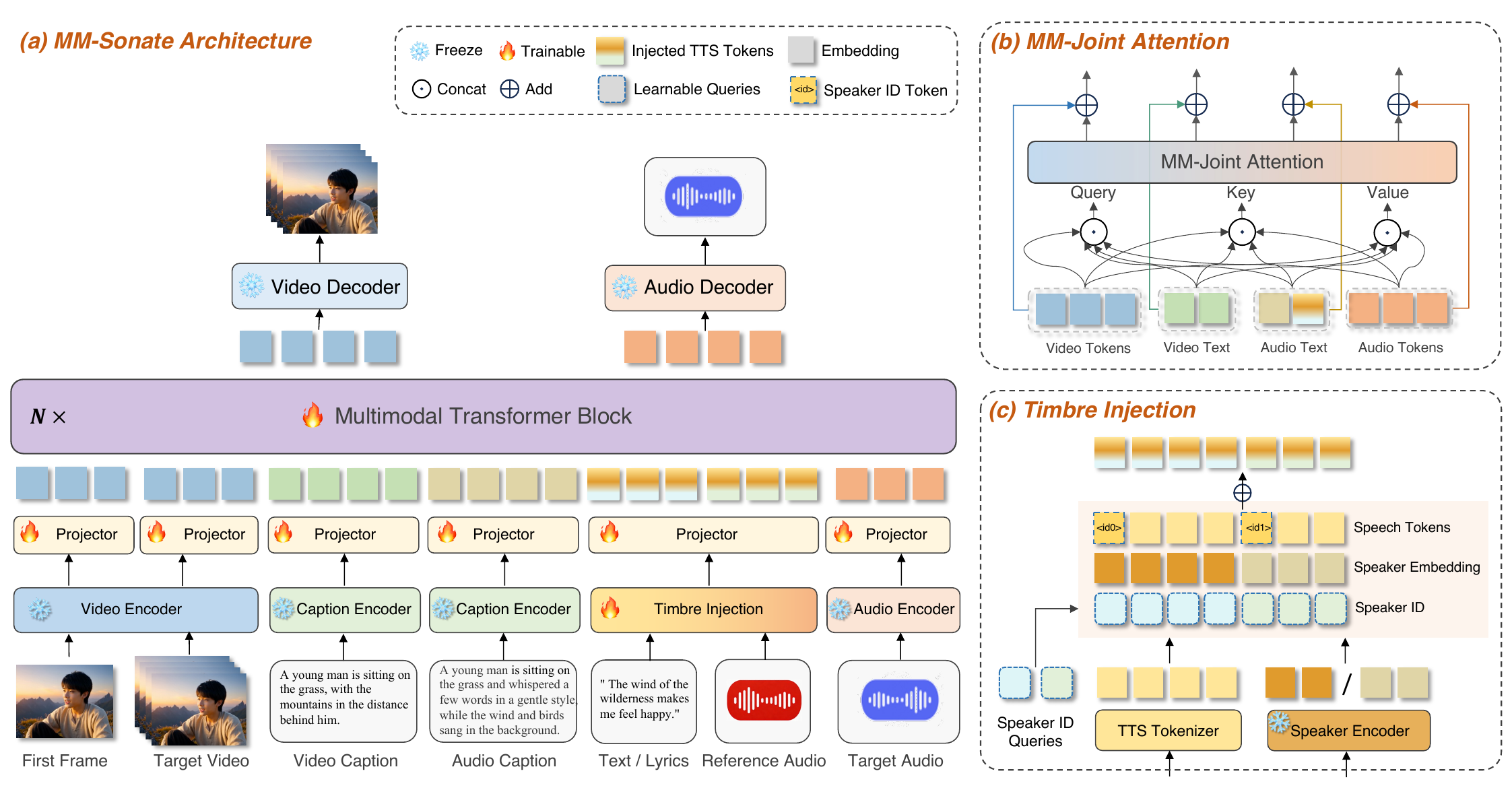}
 \caption{ The framework is a multimodal flow-matching model enabling joint audio-video generation with fine-grained control. The unified instructions combining video/audio captions with phoneme sequences for precise content alignment. The reference audio for zero-shot voice cloning. The first frame image for visual conditioning. A dedicated mechanism injects speaker embeddings (from reference audio) and ID embeddings (for multi-speaker control) directly into the phoneme sequence via element-wise addition. These multimodal features are then fused and processed by the MM-DiT backbone to model the joint distribution of audio and video latents.}
 \label{fig:model}
\end{figure*}

\subsection{Model Architecture Overview}
As illustrated in Figure~\ref{fig:model}, MM-Sonate introduces a multimodal controllable framework for joint audio-video generation with zero-shot timbre cloning capabilities.
During training, the instruction encoder, speaker encoder, video encoder, video decoder, audio encoder, and audio decoder remain frozen as pre-trained modules. The first-layer joint diffusion transformer takes temporally concatenated instruction embeddings and phoneme embeddings as text-modal conditioning inputs, while receiving noised mel-spectrogram VAE latents as audio-modal inputs and noised video VAE latents as video-modal inputs. Beyond generating video content from natural language descriptions, the model enables multi-attribute control in speech scenarios through natural language instructions, encompassing gender, age, emotion, style, and accent attributes, while supporting multi-speaker dialogue generation. Additionally, the framework supports zero-shot timbre cloning for arbitrary speakers using reference audio inputs.
For music generation scenarios, the model similarly achieves natural language control over singer characteristics (e.g., gender and age), music genre, instrumentation, melody, and emotional expression, while enabling singer voice cloning through reference audio. In sound effect generation, the model produces controlled outputs via natural language instructions.
\subsection{Unified Instruction-Guided Input}
To enable unified audio-video generation, we design a standardized instruction-phoneme input format that maintains consistent text modality inputs across all tasks, as illustrated in Fig. \ref{fig:model}. This format comprises three key components: a video instruction description, an audio instruction description, and a phoneme sequence. The video instruction description encapsulates the complete visual scene information required for generation.
The audio instruction description specifies the desired acoustic attributes of the output\cite{qiang2025instructaudio}. For speech synthesis, these attributes include speaker characteristics such as gender, age, emotion, style, and accent. In dialogue scenarios, we provide separate descriptions for each of the two speakers and prepend special tokens [S0] and [S1] to their respective text inputs to differentiate the utterances. Similarly, for music generation, the description defines attributes like the singer's timbre (e.g., gender and age), music genre, instrumentation, melody, and emotional expression. For sound effect generation, the audio description contains the relevant sonic scene information. In both speech and music tasks, the input text (for speech) or lyrics (for music) is converted into a phoneme sequence using a Grapheme-to-Phoneme (G2P) model \cite{qiang2022back}.
This unified representation allows a single model to seamlessly process text modality inputs for diverse generation tasks. It is important to note that in the T2VA and TI2VA tasks, the timbre of the generated audio is solely controlled by the audio instruction description. However, in the TA2VA and TIA2VA tasks, the timbre is derived from the input reference audio. These multimodal conditional inputs can coexist; for instance, a reference audio input can be accompanied by an audio instruction description such as "this woman speaks happily and surprisingly...".

\subsection{Voice Cloning and Multi-Speaker control}
\subsubsection{Timbre Injection}
To support both timbre cloning tasks (TA2VA, TIA2VA) and non-timbre cloning tasks (T2VA, TI2VA), we design a simple yet effective timbre injection mechanism. We pre-train a speaker encoder on a speaker classification task to extract fixed-length global speaker embeddings $S \in \mathbb{R}^{B \times D}$. Given input phoneme vectors $P \in \mathbb{R}^{B \times L \times D}$, we replicate and upsample $S$ to match the sequence length of $P$, yielding $S \in \mathbb{R}^{B \times L \times D}$. Timbre information is then injected through element-wise addition: $P = P + S \in \mathbb{R}^{B \times L \times D}$. For T2VA and TI2VA tasks without timbre cloning, the input speaker embedding is replaced with a zero vector of identical shape.

\subsubsection{Multi-Speaker Dialogue Control}
To ensure stable control in multi-speaker dialogue scenarios, we additionally introduce learnable speaker-specific ID embeddings. In a two-speaker dialogue scenario, the input phoneme vectors for the two utterances are denoted as $P_0 \in \mathbb{R}^{B \times L_0 \times D}$ and $P_1 \in \mathbb{R}^{B \times L_1 \times D}$, respectively. Two reference audio samples are provided as timbre inputs, from which we extract speaker embeddings and upsample them according to their corresponding phoneme sequence lengths: $S_0 \in \mathbb{R}^{B \times L_0 \times D}$ and $S_1 \in \mathbb{R}^{B \times L_1 \times D}$. Simultaneously, we map the speaker count to two learnable ID embeddings, $I_0 \in \mathbb{R}^{B \times D}$ and $I_1 \in \mathbb{R}^{B \times D}$, which are upsampled to match their respective phoneme sequence lengths: $I_0 \in \mathbb{R}^{B \times L_0 \times D}$ and $I_1 \in \mathbb{R}^{B \times L_1 \times D}$. The final phoneme representations are obtained by summing the three components: $P_0 = P_0 + S_0 + I_0 \in \mathbb{R}^{B \times L_0 \times D}$ and $P_1 = P_1 + S_1 + I_1 \in \mathbb{R}^{B \times L_1 \times D}$. These ID embeddings effectively improve control stability in multi-speaker dialogue scenarios, even for T2VA and TI2VA tasks without timbre cloning.

\subsection{Flexible Multi-Task Training Strategy} 
To empower the model with omni-modal generation capabilities within a unified framework, we employ a stochastic modality masking strategy during training. Instead of partitioning the training data into separate task-specific subsets, we formulate the optimization objective as learning the conditional distribution $p(V, A | C_{active})$, where $C_{active}$ represents a dynamic subset of condition embeddings (i.e., $C_{active} \subseteq \{c_{\text{text}}, c_{\text{image}}, c_{\text{audio}}\}$).

Specifically, for each training sample containing the full tuple of (text, reference image and reference audio), we independently apply Boolean masking to the conditioning inputs based on pre-defined probabilities. When a specific input modality is masked, its corresponding embedding sequence is replaced by a learnable placeholder token. This mechanism enables the model to effectively simulate diverse inference scenarios within a single training pass: T2VA is simulated by masking both reference image and audio inputs; TI2VA and TA2VA are achieved by selectively masking the reference audio or image, respectively; and TIA2VA involves no masking, utilizing the full set of conditions. Furthermore, to support the CFG strategy during inference, we randomly drop all conditional inputs with a small probability (e.g., 10\%) to approximate the unconditional distribution.

\subsection{Noise-Based Negative Conditioning} 
\label{section:cfg_inference}

CFG has become a de facto standard in generative modeling for enhancing sample fidelity and adherence to conditions. In the context of audio-video generation, the standard CFG formulation extrapolates the predicted noise $\epsilon$ between a conditional estimate $\epsilon_{cond}$ and an unconditional estimate $\epsilon_{uncond}$:
\begin{equation}
    \tilde{\epsilon} = \epsilon_{uncond} + w \cdot (\epsilon_{cond} - \epsilon_{uncond})
\end{equation}
where $w$ is the guidance scale. Traditionally, the unconditional branch $\epsilon_{uncond}$ is constructed by replacing the text prompt with a null string or a negative prompt. However, MM-Sonate introduces a complex acoustic conditioning mechanism via the speaker encoder, which necessitates a multimodal approach to negative prompting that extends beyond the textual domain.

A naive approach to constructing the unconditional acoustic branch is to use a zero vector (i.e., $S = 0$) as the negative speaker embedding. However, in our training strategy, a zero vector specifically signals the T2VA task, where the model is trained to generate a random or context-appropriate voice from text instructions alone. Consequently, using a zero vector as the negative condition in CFG merely instructs the model to diverge from the "random voice generation" mode rather than steering it away from low-quality audio.
To address this, we propose a Noise-based Negative Speaker Embedding strategy. We construct the negative speaker embedding $S_{neg}$ by feeding noise into the pre-trained speaker encoder, establishing a repulsive force against unstructured acoustic features.

To systematically identify the optimal negative audio prior, we evaluated eight distinct strategies: a zero-vector baseline, Gaussian white noise at six progressive energy levels, and naturally collected white noise.

\section{Data Preparation}
\label{section:dataset}

To achieve both high-fidelity zero-shot voice cloning and robust general-domain generation, we curate two distinct datasets: a specialized synthetic dataset constructed via a strict verification pipeline, and a massive-scale proprietary corpus for multimodal pre-training.

\subsection{High-Fidelity Synthetic Timbre Dataset}
\label{subsec:synthetic_dataset}

\begin{figure*}[t]
 \centering
 \includegraphics[width=0.7\linewidth]{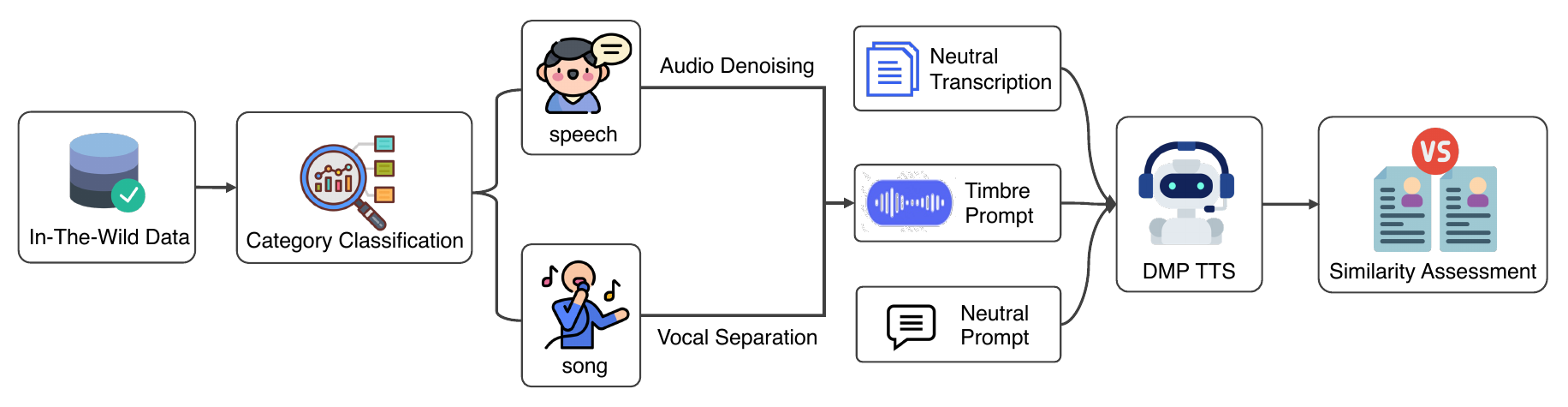}
 \caption{\textbf{The data construction pipeline for the high-fidelity synthetic timbre dataset.} The process begins with extracting clean acoustic prompts from raw speech and music via denoising and vocal separation tools. These prompts condition DMP-TTS~\cite{yin2025dmp} to synthesize speech from neutral texts. Finally, a WavLM-based\cite{chen2022wavlm} speaker verification module filters out samples with low cosine similarity to ensure high speaker identity preservation.}
 \label{fig:data_pipeline}
\end{figure*}

Constructing a dataset for zero-shot voice cloning requires decoupling speaker timbre from other acoustic factors (e.g., emotion, background noise) while maintaining high identity similarity. As illustrated in Figure~\ref{fig:data_pipeline}, we implemented a rigorous synthetic data construction pipeline using DMP-TTS~\cite{yin2025dmp}, a zero-shot text-to-speech model. 

First, to obtain high-quality acoustic prompts, we sampled raw speech and music data from our internal repository. We utilized proprietary source separation tools to denoise raw speech and extract dry vocals from musical tracks, ensuring the reference timbre was free from environmental artifacts. These processed segments served as the acoustic conditions. 
Second, to eliminate emotional leakage and ensure stability, we synthesized speech based on a pre-defined set of emotionally neutral texts, explicitly instructing the TTS model to generate neutral-style speech. 
Finally, to guarantee the preservation of speaker identity, we employed a speaker verification filter based on WavLM-large~\cite{chen2022wavlm}. We calculated the cosine similarity between the speaker embeddings of the synthesized speech and the ground truth prompts, discarding any samples that failed to meet a strict similarity threshold. This process yielded a high-quality dataset specifically optimized for learning timbre representations.

\subsection{Large-Scale Multimodal Pre-training Corpus}
\label{subsec:pretraining_corpus}

For broad-domain capability, we utilize a large-scale proprietary dataset comprising approximately 100 million aligned samples. Each sample consists of a quadruplet: video, audio, caption, and transcript (speech text or music lyrics). All data annotations were generated or verified using proprietary automated tools.
The corpus spans a diverse range of auditory domains, including human speech, environmental sound effects, and musical compositions. To support high-resolution generation, the samples range from 3 to 15 seconds in duration, with audio sampling rates between 16 kHz and 44.1 kHz, and video resolutions varying from 240p to 720p. This massive-scale data provides the fundamental world knowledge and multimodal alignment priors required for the model.

\section{Experiments}

\begin{table*}[h]
\setlength\tabcolsep{6pt}
\centering
\renewcommand{\arraystretch}{1.2}
\caption{Main comparisons of audio-video joint generation.}
\vspace{-7pt}
\label{tab:main-table}
\resizebox{.95\linewidth}{!}{
\begin{tabular}{@{}l|c|ccc|cccc|ccc@{}}
\toprule
\multirow{2}{*}{\textbf{Method}} & \multirow{2}{*}{\textbf{Framework}} & \multicolumn{3}{c|}{\textbf{Video}} & \multicolumn{4}{c|}{\textbf{Audio}} & \multicolumn{3}{c}{\textbf{AV Consistency}} \\ \cmidrule(lr){3-5} \cmidrule(lr){6-9} \cmidrule(lr){10-12} 
 &  & \textbf{MS}$\uparrow$ & \textbf{AS}$\uparrow$ & \textbf{ID}$\uparrow$ & \textbf{FD}$\downarrow$ & \textbf{KL}$\downarrow$ & \textbf{CLAP}$\uparrow$ & \textbf{WER}$\downarrow$ & \textbf{AV-A}$\downarrow$ & \textbf{SNC}$\uparrow$ & \textbf{IB}$\uparrow$ \\ 
\midrule
AudioLDM+TemoTkn\cite{liu2023audioldm, yariv2024diverse} & \multirow{3}{*}{Cascaded} & 0.05 & 0.12 & 0.07 & 2.05 & 2.53 & 0.229 & -- & 1.15 & 2.68 & 0.137 \\ 
OpenSora+FoleyGen\cite{openaisora2024, mei2024foleygen} &  & 0.42 & 0.44 & 0.56 & 3.69 & 3.08 & 0.212 & -- & 0.92 & 2.76 & 0.159 \\
OpenSora+See\&Hear\cite{openaisora2024, Xing2024SeeingAH} &  & 0.42 & 0.44 & 0.56 & 2.26 & 2.97 & 0.206 & 0.337 & 0.86 & 2.85 & 0.164 \\ 
\midrule
JavisDiT\cite{liu2025javisdit} & \multirow{5}{*}{Unified} & 0.18 & 0.36 & 0.22 & 1.95 & 5.17 & 0.228 & 0.256 & 0.92 & 3.94 & 0.231 \\
SVG\cite{ishii2024simple} &  & 0.40 & 0.41 & 0.25 & 1.55 & 3.62 & 0.080 & -- & \textbf{0.72} & 4.07 & 0.206 \\
Universe-1\cite{wang2025universe} &  & 0.20 & 0.47 & 0.25 & 1.55 & 1.25 & 0.160 & 0.180 & 0.98 & 3.92 & 0.198 \\
Ovi\cite{low2025ovi} &  & \textbf{0.58} & 0.48 & 0.46 & 1.50 & 1.19 & 0.224 & 0.035 & 0.82 & 4.28 & 0.214 \\ 
\rowcolor{gray!10}
MM-Sonate &  & 0.48 & \textbf{0.56} & \textbf{0.59} & \textbf{1.43} & \textbf{1.16} & \textbf{0.242} & \textbf{0.020} & 0.76 & \textbf{6.51} & \textbf{0.27} \\ 
\bottomrule
\end{tabular}
}
\end{table*}
\subsection{Implementation Details}
MM-Sonate is built with 20 billion parameters and utilizes a flow matching feedforward dimension of 4096. Its backbone consists of 32 joint diffusion transformer layers equipped with RoPE positional encoding \cite{su2024roformer}. For text conditioning, we adopt a phoneme encoder with a feedforward dimension of 512, alongside Qwen2.5-7B \cite{qwen2025qwen25technicalreport} serving as the caption encoder. The video VAE accommodates inputs of varying resolutions and frame rates, applying 16× spatial compression (height and width) to produce embeddings at a temporal rate of 3 Hz. Furthermore, audio processing is handled by a mel encoder that downsamples 44.1 kHz waveforms by a factor of 1024, yielding embeddings at 43 Hz. 

\subsection{Compared Method}
To comprehensively evaluate MM-Sonate, we benchmark against SOTA models across three distinct tasks. Notably, since existing T2VA models do not support timbre cloning capabilities, we compare MM-Sonate with TTS models. It is important to note that all TTS and TTM comparisons for MM-Sonate simultaneously generate accompanying video content. For audio-video joint generation, we compare against both cascade systems, including AudioLDM+TemoTkn \cite{liu2023audioldm, yariv2024diverse}, OpenSora+FoleyGen \cite{openaisora2024, mei2024foleygen} and OpenSora+See\&Hear \cite{openaisora2024, Xing2024SeeingAH}, and unified models, such as JavisDiT \cite{liu2025javisdit}, SVG \cite{ishii2024simple}, Universe-1 \cite{wang2025universe}, and the current open-source state-of-the-art model Ovi \cite{low2025ovi}. For TTS task, we compare against MaskGCT \cite{wang2024maskgct}, E2-TTS \cite{eskimez2024e2}, F5-TTS \cite{chen2024f5}, ZipVoice \cite{zhu2025zipvoice}, M3-TTS \cite{wang2025m3}, and CosyVoice2 \cite{du2024cosyvoice2}. For TTM task, we compare against DiffRhythm+\cite{chen2025diffrhythm+} and ACE-Step\cite{gong2025ace}. All evaluations are conducted on VerseBench, the benchmark dataset released with Universe-1 \cite{wang2025universe}.

\subsection{Evaluation Metrics}
We conducted human evaluations to assess the model's performance on speaker identity preservation, a critical capability for personalized generation. We adopt a multi-dimensional evaluation strategy covering video quality, audio fidelity, and cross-modal alignment.
\begin{itemize}[leftmargin=*]
\item Video Quality: We report Motion Score (MS) based on RAFT~\cite{teed2020raft} optical flow to quantify dynamic realism. Aesthetic Score (AS) is a composite metric derived from MANIQA~\cite{yang2022maniqa}, aesthetic-predictor-v2-5~\cite{aesthetic-predictor-v2-5}, and Musiq~\cite{ke2021musiq} to assess visual fidelity. Identity preservation is measured by ID Consistency (ID), computed as the mean DINOV3~\cite{simeoni2025dinov3} feature similarity between reference images and generated frames.

\item Audio Quality: We measure the distributional distance using Fréchet Distance (FD) and KL Divergence on mel-spectrogram features extracted via PANNs~\cite{kong2020panns}. Semantic alignment between audio and text is evaluated using the CLAP score~\cite{wu2023large}.

\item Speech \& Voice Cloning: Speech intelligibility is quantified by the Word Error Rate (WER), derived from Whisper-large-v3~\cite{radford2023robust} transcriptions. To assess speaker identity preservation, we employ specific metrics depending on the evaluation context. For the main comparative benchmark against state-of-the-art TTS models (Table~\ref{tab:tts1}), we follow the SeedTTS\cite{anastassiou2024seed} evaluation protocol and report SIM-o, calculated using the WavLM-large~\cite{chen2022wavlm} speaker verification model. In our ablation studies regarding negative conditioning strategies (Fig.~\ref{fig:energy_ranking} and \ref{fig:energy_sensitivity}), we utilize the Resemblyzer encoder\footnote{\url{https://github.com/resemble-ai/Resemblyzer}} to measure Speaker Consistency.

\item Audio-Visual Synchronization: To strictly assess temporal alignment, we report AV-A (Audio-Video Alignment) distance computed via Synchformer~\cite{iashin2024synchformer}. For lip synchronization specifically, we report the SyncNet Confidence (SNC) score~\cite{chung2016out}. Additionally, global cross-modal alignment is measured via ImageBind (IB).

\item Music Evaluation: Musical output is assessed using the SongEval\cite{yao2025songeval}, focusing on Coherence, Musicality, Memorability, Clarity, and Naturalness.
\end{itemize}

\begin{table*}[t]
\centering
\caption{Comparison of TTS models on speaker similarity and word error rate performance.}
\label{tab:tts1}
\resizebox{0.6\linewidth}{!}{ 
\begin{tabular}{l|c|cc|cc}
\toprule
\multirow{2}{*}{\textbf{Model}} & \multirow{2}{*}{\textbf{Framework}} & \multicolumn{2}{c|}{\textbf{SIM-o $\uparrow$}} & \multicolumn{2}{c}{\textbf{WER(\%) $\downarrow$}} \\
\cmidrule(lr){3-4} \cmidrule(lr){5-6}
 &  & \textbf{EN} & \textbf{ZH} & \textbf{EN} & \textbf{ZH} \\
\midrule
Ground Truth & -- & 0.734 & 0.755 & 2.14 & 1.25 \\
\midrule
MaskGCT\cite{wang2024maskgct} & \multirow{6}{*}{TTS} & \textbf{0.713} & \textbf{0.773} & 2.26 & 2.40 \\
E2-TTS\cite{eskimez2024e2} &  & 0.706 & 0.713 & 2.49 & 1.91 \\
F5-TTS\cite{chen2024f5} &  & 0.664 & 0.750 & 1.89 & 1.53 \\
ZipVoice\cite{zhu2025zipvoice} &  & 0.697 & 0.751 & 1.70 & 1.40 \\
M3-TTS\cite{wang2025m3} &  & 0.604 & 0.621 & \textbf{1.36} & \textbf{1.31} \\
CosyVoice2\cite{du2024cosyvoice2} &  & 0.652 & 0.748 & 2.57 & 1.45 \\
\midrule
\rowcolor{gray!10}
MM-Sonate & Unified & 0.604 & 0.691 & 2.065 & 2.768 \\
\bottomrule
\end{tabular}
}
\end{table*}

\begin{table*}[t]
\centering
\caption{Performance comparison of TTM models on SongEval metrics.}
\label{tab:music}
\resizebox{0.6\linewidth}{!}{ 
\begin{tabular}{l|c|ccccc}
\toprule
\multirow{2}{*}{\textbf{Model}} & \multirow{2}{*}{\textbf{Framework}} & \multicolumn{5}{c}{\textbf{SongEval $\uparrow$}} \\
\cmidrule(lr){3-7}
 &  & \textbf{Coh} & \textbf{Mus} & \textbf{Mem} & \textbf{Cla} & \textbf{Nat} \\
\midrule
Ground Truth & -- & 3.60 & 3.52 & 3.56 & 3.43 & 3.34 \\
\midrule
DiffRhythm+\cite{chen2025diffrhythm+} & \multirow{2}{*}{TTM} & 2.68 & 2.61 & 2.57 & 2.48 & 2.37 \\
ACE-Step\cite{gong2025ace} &  & 2.89 & 2.87 & 2.83 & 2.77 & 2.71 \\
\midrule
\rowcolor{gray!10}
MM-Sonate & Unified & \textbf{3.00} & \textbf{3.01} & \textbf{2.98} & \textbf{2.88} & \textbf{2.72} \\
\bottomrule
\end{tabular}
}

\begin{tablenotes} 
      \scriptsize
      \item \textbf{Note:} Coh = Coherence, Mus = Musicality, Mem = Memorability, Cla = Clarity, Nat = Naturalness.
  \end{tablenotes}
  
\end{table*}

\subsection{Results and Analysis}
\subsubsection{Ablation Studies}

\begin{figure}[t]
    \centering
    \includegraphics[width=\linewidth]{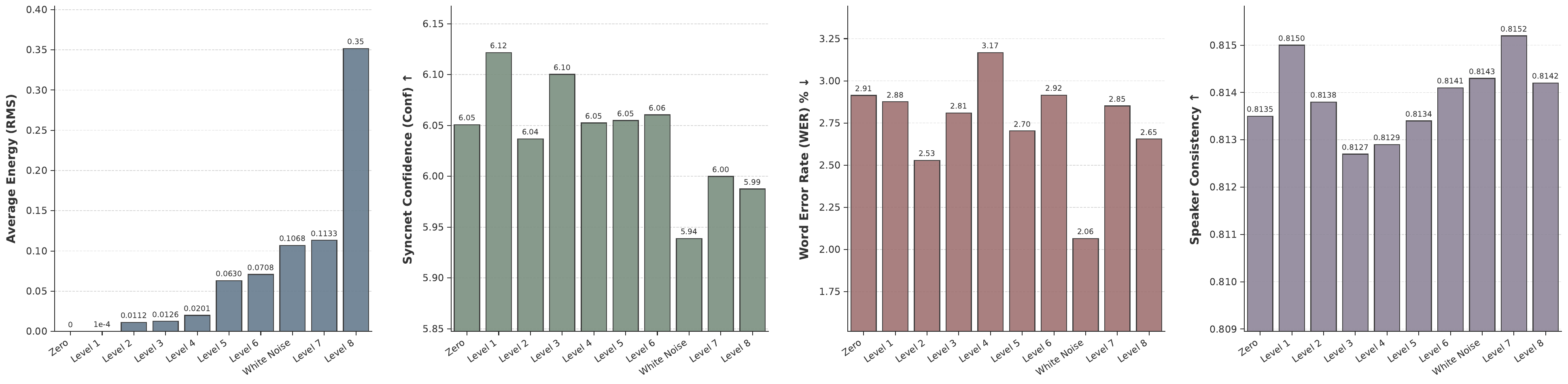}
    \caption{\textbf{Performance comparison of different negative conditioning strategies.}}
    \label{fig:energy_ranking}
\end{figure}

\begin{figure}[t]
    \centering
    \includegraphics[width=\linewidth]{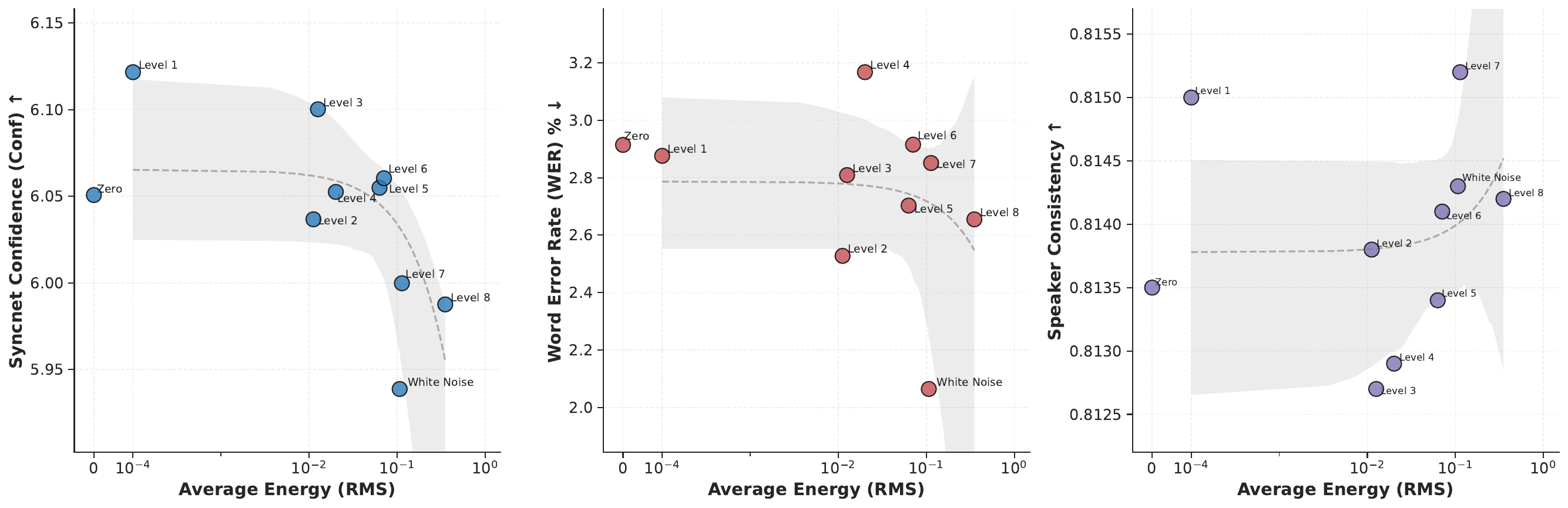}
    \caption{\textbf{Sensitivity analysis of negative embedding energy on generation metrics.}}
    \label{fig:energy_sensitivity}
\end{figure}

\textbf{Analysis of Negative Conditioning Strategies.}
As illustrated in Figure~\ref{fig:energy_ranking} and Figure\ref{fig:energy_sensitivity}, varying the energy levels of Gaussian white noise reveals a trade-off between modalities. We observed that as the noise energy increases, syncnet confidence exhibits a gradual decline, while speaker consistency shows a slight upward trend. However, these variations across Gaussian energy levels remain relatively marginal, suggesting that merely 

\textbf{Superiority of Natural Noise.}
Most notably, our experiments highlight the distinct advantage of using naturally collected white noise. As shown in Figure~\ref{fig:energy_ranking}, the natural noise strategy achieves the lowest WER among all settings, significantly outperforming the zero-vector and Gaussian variants. Specifically, compared to the worst-performing setting (WER 3.17\%), the natural noise setting achieves a WER of 2.06\%, representing a substantial relative improvement of 1.11\%.
While this strategy results in a slight compromise in syncnet confidence dropping from the optimal 6.12 to 5.94 (a decrease of only 0.18), it maintains robust speaker timbre consistency. We attribute the superiority of natural noise to the manifold hypothesis: naturally collected noise lies closer to the true distribution of the background noise present in the training data compared to synthetic Gaussian distributions or null vectors. Consequently, the speaker encoder maps natural noise to a more effective region in the embedding space, providing a more semantically meaningful "negative" reference for the flow matching process. Based on these findings, we adopt naturally collected white noise as the default negative speaker embedding for inference.

\begin{figure*}[t]
  \centering
  \includegraphics[width=0.6\linewidth]{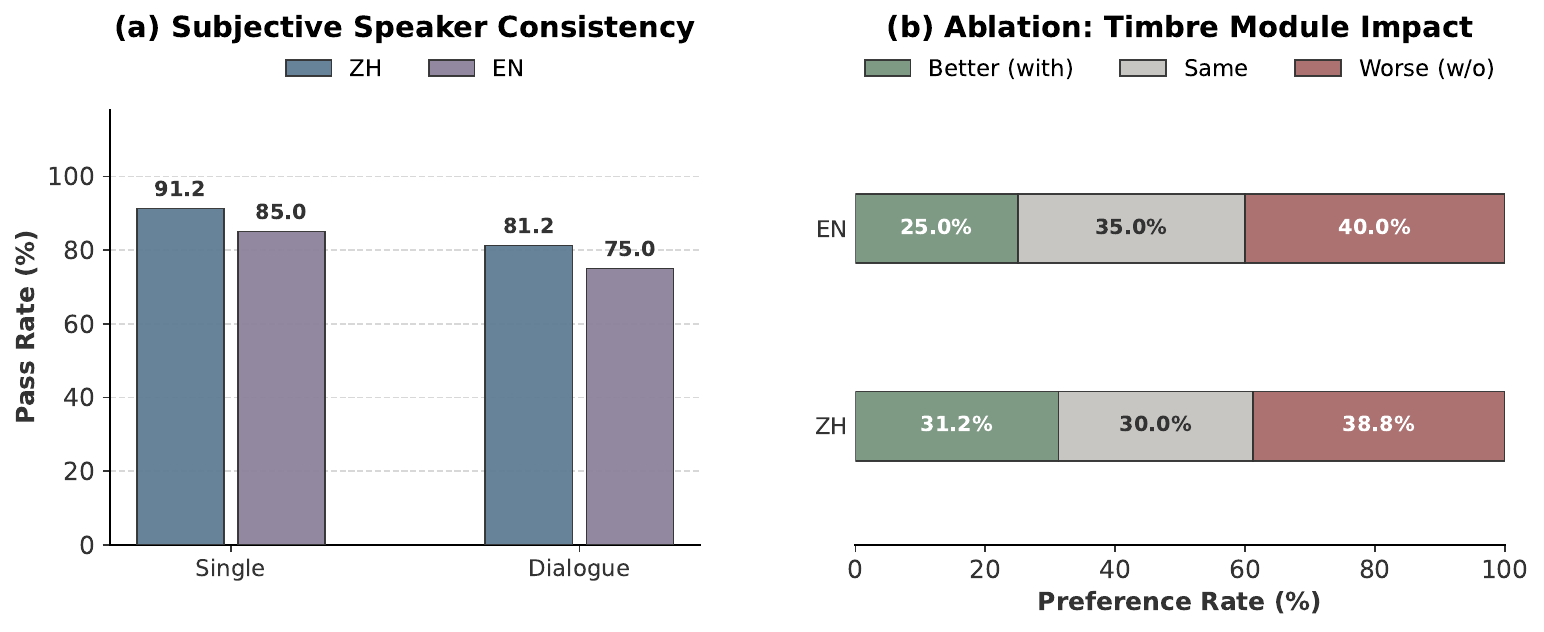} 
  \vspace{-5pt}
  \caption{
  \textbf{Subjective evaluation results.} 
  (a) Pass rates for speaker consistency, confirming robust identity preservation in both single-speaker and complex dialogue scenarios.
  (b) Ablation study comparing the full MM-Sonate against a baseline on standard generation tasks (T2VA/TI2VA). Although adding the timbre module causes a slight performance trade-off, the high "Same" and "Better" rates indicate that basic generation quality remains robust, while the model gains the unique capability for zero-shot voice cloning (TA2VA/TIA2VA).
  }
  \label{fig:subjective_stats}
\end{figure*}

\textbf{Timbre Injection Impact} Integrating the timbre injection mechanism introduces additional complexity to the conditioning space, which theoretically poses a risk of interfering with the model's standard generation capabilities (T2VA/TI2VA). To quantify this trade-off, we conducted a side-by-side preference test comparing the full MM-Sonate model (supporting reference audio) against a variant trained without the timbre injection on standard non-cloning tasks. 
As shown in Figure~\ref{fig:subjective_stats}(b), while there is a slight preference for the specialized text-only model in some cases (indicated by the "Worse" portion), the majority of samples are rated as "Same" or "Better" (GSB score: 0.89 for ZH, 0.83 for EN). 
We argue that this marginal degradation in standard tasks is an acceptable cost. Unlike the baseline, the full MM-Sonate unlocks entirely new generation paradigms—specifically TA2VA/TIA2VA, enabling zero-shot voice cloning capabilities that are otherwise impossible.

\subsubsection{Evaluation of Audio-Video Generation}
Table~\ref{tab:main-table} presents the quantitative comparison across video, audio, and consistency metrics. MM-Sonate establishes a new state-of-the-art in unified audio-video synthesis, particularly excelling in acoustic fidelity and cross-modal synchronization.

\textbf{Audio Fidelity and Intelligibility.} 
A distinct advantage of MM-Sonate lies in its ability to generate high-fidelity, linguistically accurate audio. Our model achieves the lowest FD (1.43) and KL (1.16) scores, surpassing the previous SOTA, Ovi (FD 1.50). This indicates that our latent audio codec and flow-matching prior effectively capture complex acoustic distributions better than competing unified baselines. 
Most strikingly, in terms of speech intelligibility, MM-Sonate achieves a WER of 0.020, which is an order of magnitude improvement over open-source baselines like OpenSora+See\&Hear (0.337) and JavisDiT (0.256). While existing T2AV models often struggle with mumbled or incoherent speech generation, our phoneme-guided architecture ensures the production of clear, distinct semantic content.

\textbf{Fine-Grained Synchronization.} 
Our model demonstrates a significant leap in temporal alignment capabilities. On the SyncNet Confidence (SNC) metric—a proxy for lip-synchronization quality—MM-Sonate scores 6.51, substantially outperforming the closest competitor, Ovi (4.28), and the cascaded baseline AudioLDM+TemoTkn (2.68). This sharp contrast suggests that MM-Sonate moves beyond coarse-grained semantic matching to achieve precise, frame-level alignment between lip movements and phonemes. This observation is further supported by our top-ranking ImageBind (IB) score of 0.27, confirming robust global semantic consistency.

\textbf{Visual Quality.} 
Despite the increased difficulty of modeling joint distributions, MM-Sonate maintains highly competitive visual quality. We achieve the highest Aesthetic Score (AS 0.56), tying with the strong video-centric baseline OpenSora+FoleyGen. Furthermore, MM-Sonate demonstrates robust identity preservation, surpassing cascaded approaches in Identity Consistency (ID 0.59 vs. 0.56). This indicates that our unified framework not only solves the critical challenge of intrinsic audio-visual coherence that cascaded systems lack but also delivers superior visual fidelity comparable to or better than leading baselines (e.g., Ovi MS 0.48 vs. 0.58).

\subsubsection{Evaluation of Voice Cloning and Music}

Since current unified audio-video joint generation models lack the capability for zero-shot timbre cloning, we benchmark MM-Sonate against specialized unimodal systems (TTS and TTM). Crucially, while these baselines generate only audio, MM-Sonate synthesizes synchronized video content simultaneously, tackling a significantly more complex generative task involving joint cross-modal distribution modeling.

\textbf{Zero-Shot Voice Cloning.} 
Table~\ref{tab:tts1} presents a comparison with dedicated Text-to-Speech (TTS) systems. It is crucial to note that while these baselines are specialized for audio-only synthesis, MM-Sonate tackles a significantly more complex task: generating synchronized, high-fidelity video alongside the cloned audio. 
Considering this increased complexity, the model's voice cloning fidelity remains highly competitive. On English prompts, MM-Sonate achieves a Speaker Similarity (SIM-o) of 0.604, performing on par with M3-TTS, and a SIM-o of 0.691 on Chinese prompts. Although this represents a slight trade-off compared to the top-performing audio-only model MaskGCT (0.713), we argue this is an acceptable cost for gaining full audio-visual generation capabilities.
Furthermore, MM-Sonate excels in linguistic precision, a critical aspect of multimodal generation. Its English Word Error Rate (WER) of 2.065\% is superior to strong baselines like CosyVoice2 (2.57\%) and approaches the ground truth (2.14\%). These results demonstrate that MM-Sonate effectively preserves speech intelligibility and achieves robust voice cloning within a demanding joint generation framework.

We evaluated timbre fidelity in two scenarios: (1) Single-Speaker: measuring the similarity between the cloned voice and the reference audio; and (2) Dialogue: assessing the distinctness and consistency of two separate identities within a single video. As shown in Figure~\ref{fig:subjective_stats}(a), MM-Sonate achieves high pass rates for both Chinese (ZH) and English (EN). Notably, in single-speaker cloning, the model reaches 91.3\% (ZH) and 85.0\% (EN) pass rates. In complex dialogue settings, it maintains robust identity consistency (ZH: 81.3\%, EN: 75.0\%), confirming its effectiveness in multi-speaker control.

\textbf{Music Generation.} 
In the music domain (Table~\ref{tab:music}), MM-Sonate demonstrates superior structural quality compared to specialized Text-to-Music models. We outperform ACE-Step and DiffRhythm+ across all five SongEval metrics. Notably, we achieve a Musicality score of 3.01 and Coherence of 3.00, significantly higher than ACE-Step (2.87 and 2.89, respectively). This suggests that the large-scale multimodal pre-training provides richer structural priors than audio-only training, allowing MM-Sonate to generate more melodically pleasing and coherent musical compositions alongside matched visual content.

\begin{figure*}[!t]
  \centering
  \includegraphics[width=\linewidth]{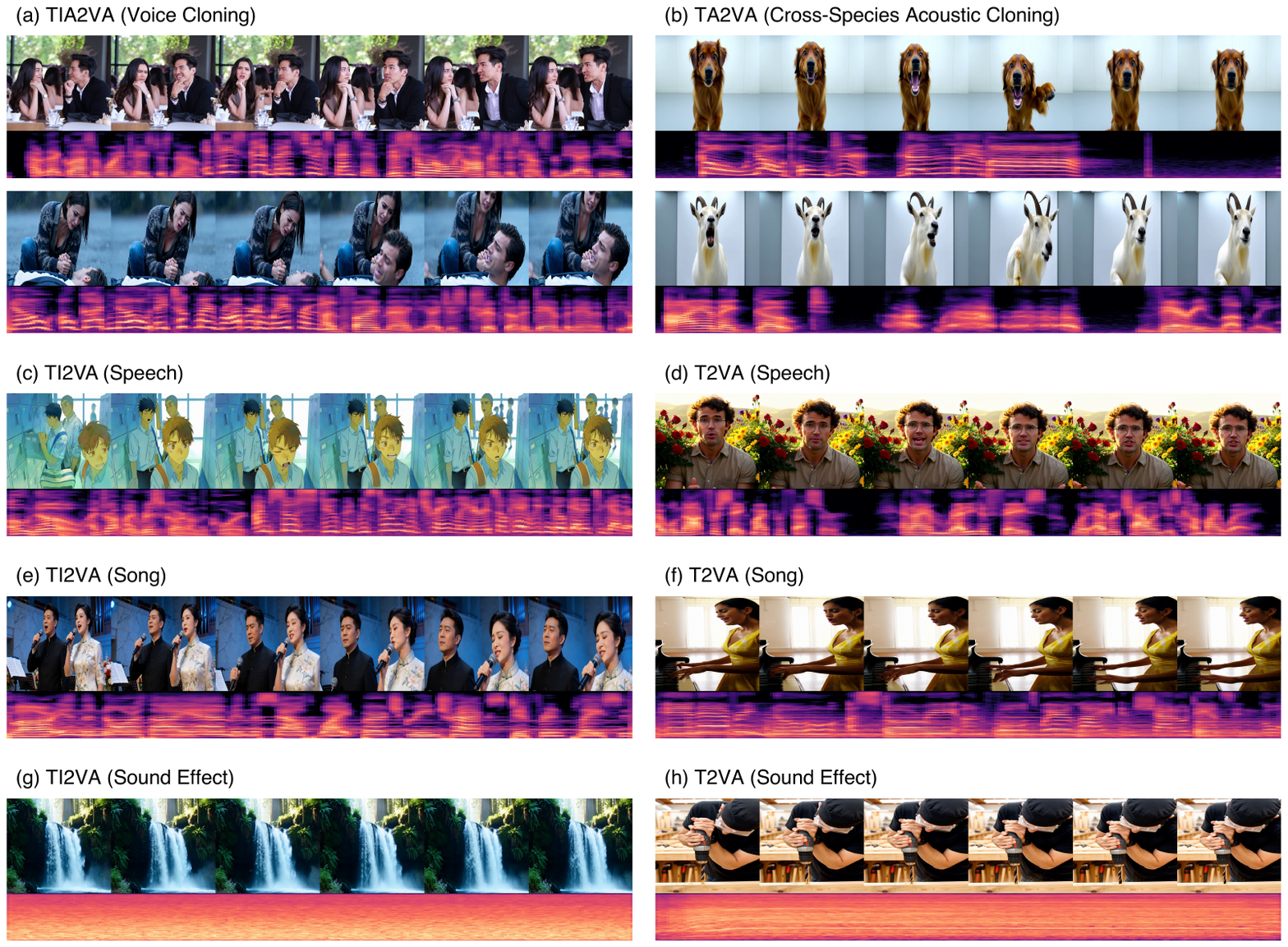}
  \caption{
  \textbf{Qualitative visualization of omni-modal audio-video generation and temporal alignment.} 
  We demonstrate MM-Sonate's versatility across diverse task configurations, ranging from zero-shot voice cloning to musical performance and sound effect generation.
  Each panel displays generated video frames (top) aligned with their corresponding mel-spectrograms (bottom), where vertical dashed lines indicate synchronization between visual dynamics and acoustic events.
  (a)-(b) Voice \& Acoustic Cloning: High-fidelity audio synthesis conditioned on reference audio. Notably, panel (b) demonstrates cross-species generalization, where the model successfully clones non-speech acoustic identities (e.g., specific dog barks and goat bleats) and synthesizes the corresponding animal vocalization movements.
  (c)-(d) Speech Synthesis: Standard text-to-speech video generation ensuring precise lip-phoneme synchronization.
  (e)-(f) Song Generation: Musical performance synthesis demonstrating rhythmic alignment for singing vocals.
  (g)-(h) Sound Effect: Generation of environmental sounds synchronized with visual motion.
  Across all scenarios—whether generating from scratch (T2VA/TA2VA) or animating static images (TI2VA/TIA2VA)—the model maintains strict semantic consistency and temporal coherence.
  }
  \label{fig:qualitative_omni}
\end{figure*}

\subsubsection{Qualitative Analysis}
To visually substantiate the quantitative metrics reported above, we present a frame-level analysis of generated samples across diverse task configurations in Figure~\ref{fig:qualitative_omni}. The figure illustrates MM-Sonate's "omni-modal" capability, handling inputs ranging from text-only instructions (T2VA) to complex combinations involving reference images and audio (TIA2VA).
Crucially, the time-aligned mel-spectrograms reveal strict synchronization between visual dynamics and acoustic events.
In speech and voice cloning scenarios (Panels a-d), we observe precise alignment between lip closures and plosive sounds, as well as distinct mouth openings corresponding to vowel peaks.
Remarkably, we observe that our timbre injection mechanism generalizes beyond human speech.
As shown in Panel (b), when conditioned on animal vocalizations (e.g., a dog's bark or a goat's bleat), MM-Sonate performs cross-species acoustic cloning. It captures the specific textural identity of the reference animal sound and synthesizes a corresponding video where the animal's jaw movements and breathing patterns physically match the acoustic impulses. This suggests that the model has learned a generalized representation of "sound source identity" rather than being limited to human phonetics.
For musical generation (Panels e-f), the model accurately captures the rhythmic structure, synchronizing sustained vocal notes with prolonged visual articulations.
Furthermore, in sound effect generation (Panels g-h), high-energy spectral bursts coincide perfectly with physical motion events.
These qualitative results confirm that MM-Sonate maintains high-fidelity cross-modal coherence regardless of the conditioning modality or generative domain.

\section{Limitations, Ethics and Safety}
\label{section:limits_ethics}

While MM-Sonate demonstrates state-of-the-art capabilities in audio-video joint generation and zero-shot voice cloning, we acknowledge several limitations and ethical responsibilities inherent to this technology.

\textbf{Technical Limitations.} 
First, despite the effectiveness of our joint diffusion architecture, generating long-form content remains a challenge. Currently, the model is optimized for clips ranging from 3 to 15 seconds; maintaining semantic consistency and temporal coherence for durations exceeding this range requires further investigation into long-context attention mechanisms and autoregressive extension strategies. 
Second, although our video VAE and audio mel-encoder achieve high compression rates, this can occasionally result in the loss of high-frequency details, such as subtle texture artifacts in video or transient clarity in complex musical compositions. 
Third, while our model excels at lip synchronization for frontal and near-frontal views, extreme head poses, rapid camera movements, or heavily occluded faces can still lead to desynchronization or visual artifacts in the mouth region. Finally, the zero-shot voice cloning module, while robust, may struggle to decouple background noise from the reference speaker's timbre when provided with low-quality, noisy reference audio.

\textbf{Ethical Considerations.}
The capability of MM-Sonate to perform zero-shot voice cloning and photorealistic video synthesis introduces significant ethical risks, particularly regarding the potential for misuse in creating deepfakes, impersonation, and misinformation. The ability to clone a speaker's timbre from a brief reference clip raises concerns about biometric privacy and non-consensual voice usage. Furthermore, like all large-scale generative models trained on vast datasets, MM-Sonate is susceptible to reflecting inherent biases present in the training data, potentially leading to stereotypical representations in generated gender, race, or cultural scenarios.

\textbf{Safety and Mitigation Strategies.} We employ robust safety classifiers for both input prompts (text/image/audio) and generated outputs to filter out harmful, violent, pornographic, or hateful content. For the voice cloning feature, we are exploring strict verification mechanisms to ensure that the user has the right to use the reference voice. Additionally, we deliberately constrain the generation of specific high-risk public figures. To combat misinformation, all video and audio content generated by MM-Sonate is embedded with imperceptible watermarks. These watermarks are resilient to common post-processing attacks (e.g., compression, cropping) and allow for the automated detection of AI-generated content. The model is currently released under a controlled environment (e.g., API with tiered access) rather than as open weights, allowing us to monitor usage patterns and mitigate abuse dynamically.

\section{Conclusion}
In this work, we present MM-Sonate, a unified framework built upon the MM-DiT architecture that enables high-fidelity audio-video joint generation alongside zero-shot voice cloning.  Extensive empirical evaluations across diverse benchmarks demonstrate that our approach establishes new state-of-the-art performance in joint generation tasks, particularly excelling in fine-grained lip synchronization and speech intelligibility.  Notably, MM-Sonate achieves voice cloning fidelity comparable to leading open-source TTS specialist models and demonstrates musicality superior to dedicated TTM baselines.  Furthermore, we validate the effectiveness of natural noise injection as a robust strategy for classifier-free guidance and introduce a streamlined pipeline for constructing the high-fidelity synthetic timbre dataset. By integrating precise multimodal control, MM-Sonate offers a new paradigm for the future of audio-video joint generation.

\section*{Acknowledgement}
We acknowledge the contributions from (sorted by first name): Boyuan Jiang, Chen Zhang, Feng Deng, Jiahao Wang, Jingbin He, Jingke Li, Jingru Zhao, Junjie Yan, Liang Hou, Lingyu Zou, Ming Wen, Nan Li, Peihan Li, Pengfei Wan, Teng Ma, Xiaoyu Shi, Xijuan Zeng, Xin Tao, Xu Li, Yan Zhou, Yiran Wang, Yu Zhao, Yuan Gao, Yun Li, Yushen Chen, Yuzhe Liang, Zewen Song, Zhongliang Liu, Zihan Li, Zihao Ji, Ziyang Yuan, and Ziyu Zhang.

\bibliography{iclr2025_conference}
\bibliographystyle{iclr2025_conference}

\end{document}